# Room temperature deformation in the Fe₇Mo₆ µ-Phase

S. Schröders[1], S. Sandlöbes[1], C. Birke[1], M. Loeck[1], L. Peters[3], C.Tromas[2], S. Korte-Kerzel[1,*]

*Corresponding author: korte-kerzel@imm.rwth-aachen.de

[1]Institut für Metallkunde und Metallphysik, RWTH Aachen University, Kopernikusstraße 14, 52074 Aachen

[2]PPrime Institute, UPR 3346 CNRS –Université de Poitiers – ISAE-ENSMA, SP2MI, 11 Boulevard Marie et Pierre Curie, 86962 Futuroscope Chasseneuil Cedex

[3]Institut für Kristallographie, RWTH Aachen University, Jägerstraße 17-19, 52066 Aachen

Abstract

The role of topologically close packed (TCP) phases in deformation of superalloys and steels is still not fully resolved. In particular, the intrinsic deformation mechanisms of these phases are largely unknown including the active slip systems in most of these complex crystal structures. Here, we present a first detailed investigation of the mechanical properties of the Fe₇Mo₆ µ-phase at room temperature using microcompression and nanoindentation with statistical EBSD-assisted slip trace analysis and TEM imaging. Slip occurs predominantly on the basal and prismatic planes, resulting also in decohesion on prismatic planes with high defect density. The correlation of the deformation structures and measured hardness reveals pronounced hardening where interaction of slip planes occurs and prevalent deformation at pre-existing defects.

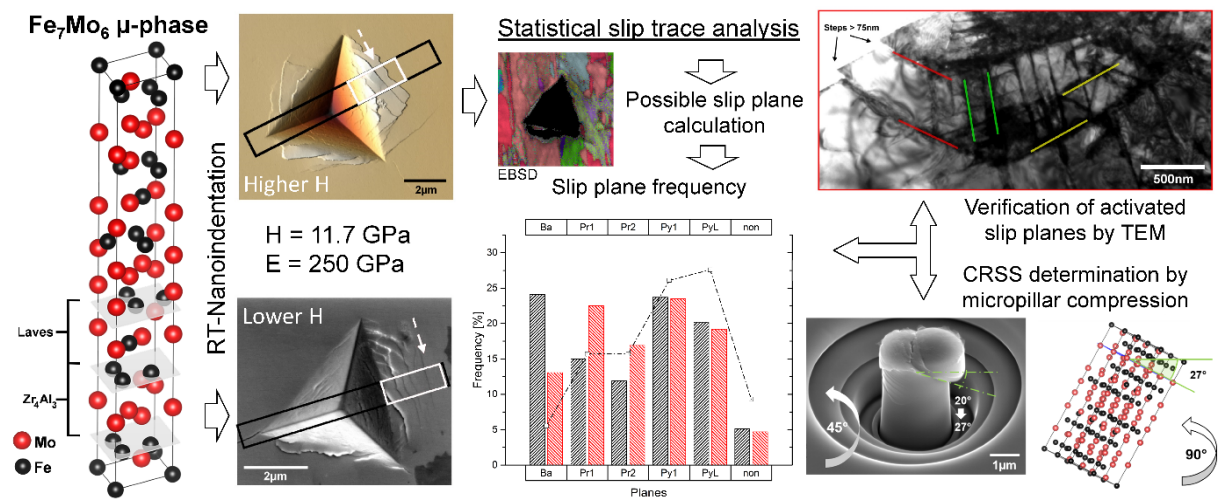

## 1  Introduction

Topologically close packed (TCP) phases, such as the µ-phase, form as precipitates in a wide range of materials. They are highly ordered phases both in terms of chemistry and stacking, although a wide range of compositions may be found (Cheng et al., 2012; Huhn et al., 2014). Their complex crystal structure usually results in pronounced brittleness of these phases (Sauthoff, 1995). The most prominent cases, where TCP phases form as precipitates, occur in heavily alloyed nickel- and iron-based superalloys, where TCP phases form either in the γ-γ'-matrix or the secondary reaction zone underneath a thermal barrier coating (Chen et al., 1980; Darolia et al., 1988; Rae and Reed, 2001; Walston et al., 1996; Yulin and Yunrong, 1982). TCP phases are further observed during joining, particularly in the presence of refractory metals such as in steel-tungsten joints implemented in fusion reactors (Matějíček et al., 2015).

Although it is commonly assumed that the µ-phase, like the other TCP phases, is detrimental to the mechanical properties and oxidation behaviour of superalloys, the underlying mechanisms are not yet fully understood and controversially discussed (Chen et al., 1980; Cheng et al., 2011; Epishin et al., 2010; He et al., 2005; Moverare et al., 2009; Pessah et al., 1992; Qin et al., 2009; Simonetti and Caron, 1998; Sims et al., 1987; Sugui et al., 2010; Tawancy, 1996; Wang et al., 2010; Wang et al., 2014; Yang et al., 2006; Zhao and Dong, 2012). In particular, it is not clear whether a certain amount of µ-phase precipitates might even be tolerated by the material without significant negative effects (Pessah et al., 1992; Simonetti and Caron, 1998).

In superalloys, the TCP µ-phase causes depletion of the γ-matrix in the solution strengthening refractory metal elements (Pessah et al., 1992; Simonetti and Caron, 1998; Sims et al., 1987; Yang et al., 2006). Dislocation pile-ups at intermetallic precipitates have been reported to cause microcracking or decohesion, embrittlement and eventually crack initiation (Chen et al., 1980; Cheng et al., 2011; Pessah et al., 1992; Qin et al., 2009; Simonetti and Caron, 1998; Tawancy, 1996; Zhao and Dong, 2012). Further, TCP phase precipitates at grain boundaries are thought to cause high stress concentrations (Pyczak et al., 2000; Sugui et al., 2010; Zhao and Dong, 2012) resulting from the high hardness of the µ-phase compared with the surrounding matrix but also the precipitates' shape and orientation relationship to the matrix. Spherical precipitates have been reported to be prone to decohesion, while needle- or plate-like precipitates have been found to lead to dislocation pile-ups and cracking (Pessah et al., 1992; Simonetti and Caron, 1998; Zhao and Dong, 2012). On the other hand, this can be inhibited due to good interfacial strength towards the matrix or a γ'-film enclosing µ-phase particles (Simonetti and Caron, 1998; Wang et al., 2010; Yang et al., 2006). During creep of γ-γ'-superalloys, µ and other TCP phase precipitates have been observed to accelerate rafting (Zhao and Dong, 2012) and to distort the unrafted (Pyczak et al., 2000; Yang et al., 2007) and, more severely, rafted microstructures (Han et al., 2008; le Graverend et al., 2011; Moverare et al., 2009; Simonetti and Caron, 1998; Sugui et al., 2010).

However, while in most studies the µ-phase precipitates are assumed not to deform themselves, ductility of needle shaped precipitates has indeed been found (Qin et al., 2009; Yang et al., 2006) and no detrimental effect on the tensile strength was found during high temperature tensile testing (Chen et al., 1980; Tawancy, 1996).

Particularly the mechanisms by which µ-phase precipitates affect creep at high temperatures are still under discussion (Cheng et al., 2011; Epishin et al., 2010; Han et al., 2008; He et al., 2005; Moverare et al., 2009; Simonetti and Caron, 1998; Sugui et al., 2010; Volek and Singer, 2004; Wang et al., 2010; Yang et al., 2006). Le Graverend et al. (le Graverend et al., 2011) have conducted FEM calculations of nickel-based superalloys considering plasticity of µ-phase precipitates at higher temperatures. However, they (le Graverend et al., 2011) were not able to exactly replicate the crystal rotations of the γ'-rafts and µ-phase needles evolving from the local stress field of matrix and precipitates. Their simulations show indeed an influence in the microstructure evolution when considering plasticity and anisotropy of µ-phase precipitates. Given the lack of mechanical data on individual slip systems, the authors therefore have assumed deformation driven by stacking fault and twin formation observed elsewhere (Carvalho et al., 2000; Qin et al., 2009; Stenberg and Andersson, 1979; Tawancy, 1996). However, no quantitative information regarding critical resolved shear stresses (CRSS) on a given system are available from these studies. The anisotropy assumed in the constitutive model may therefore be incomplete or incorrect and lead to the deviation of the simulated microstructure evolution from experimental results reported (le Graverend et al., 2011).

Therefore, knowledge of the active slip systems, the underlying dislocation mechanisms and their critical resolved shear stresses is required to understand plasticity in TCP phases and other complex crystals as well as their effect on the co-deformation in alloys with intermetallic reinforcement. The deformation behaviour of the µ-phase is still largely uninvestigated (le Graverend et al., 2011), mainly due to its pronounced brittleness at ambient temperatures. The complexity of the µ-phase's crystal structure impedes a direct identification by estimates from the unit cell, for which not even deformation of the individual building blocks has been fully understood (Chisholm et al., 2005; Krämer and Schulze, 1968; Livingston and Hall, 1990).

The hexagonal unit cell of the µ-phase is given in Figure 1a (Momma and Izumi, 2011) using the structural model by Lejaeghere et al. (Lejaeghere et al., 2011). It exhibits a complex crystallographic structure built up from Frank-Kasper polyhedra (Frank and Kasper, 1958, 1959) of CN12, CN15, CN16 and CN15 packing (Kumar et al., 1998; Wagner et al., 1995), resulting in an alternating stacking of $Zr_4Al_3$ and Laves-phase triple layers. The space group of the µ-phase is 166, $R\overline{3}m$, with 13 (or 39 in hexagonal notation) atoms per unit cell (Cieslak et al., 2014; Krishna et al., 2013; Rae et al., 2000). Its typical stoichiometry is $A_7B_6$ and the archetype is $Fe_7W_6$ (Forsyth and D'Alte da Veiga, 1962; Magneli and Westgren, 1938). Smaller A atoms are positioned in CN12 polyhedra whilst larger B atoms occupy the higher coordinated ones (Cieslak et al., 2014; Kumar et al., 1998). Within a wide range of binary and ternary alloys of early and late transition metals (Carvalho et al., 2000; Forsyth and D'Alte da Veiga, 1962; Huhn et al., 2014; Matějíček et al., 2015; Ren et al., 2014; Wagner et al., 1995), the structural parameter **a** extends between 4.723-4.769 Å and **c** between 25.48 - 25.83 Å (Magneli and Westgren, 1938; Rae et al., 2000; Stenberg and Andersson, 1979).

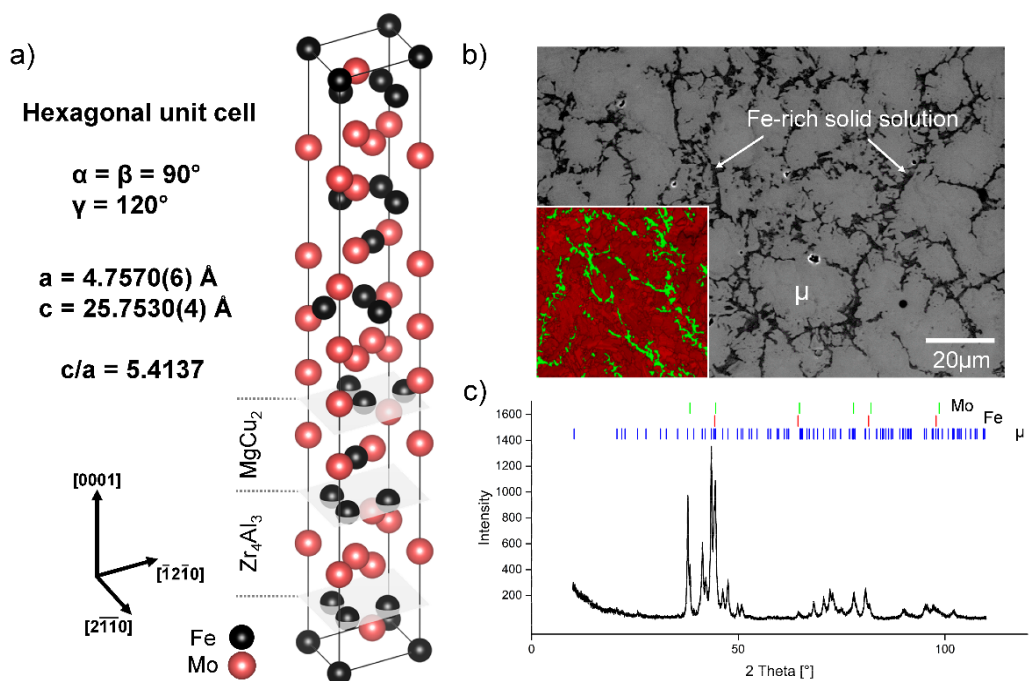

**Figure 1: Unit cell of the Fe₇Mo₆ μ-phase consisting of building blocks of Zr₄Al₃ and Laves-phase (a); (b) Secondary electron micrograph and (c) X-ray diffractogram of the heat-treated Fe-Mo 55 wt.-% alloy used within this study. The inset in (b) shows a phase mapping with overlaid image quality in greyscale indicating only iron (green) and μ-phase (red).**

Only a few recently published papers focus on the mechanical properties of the μ-phase itself, either examined by nanoindentation of precipitates within a creep deformed nickel-base superalloy (Rehman et al., 2015) and Fe-W composites (Matějíček et al., 2015), or by ab-initio assessment (Huhn et al., 2014; Ren et al., 2014) of different alloy systems. However, attempts have been made neither to identify the experimentally observed slip systems, nor to determine the CRSS of these. The defects described so far (Carvalho and De Hosson, 2001; Carvalho et al., 2000; Hiraga et al., 1983; Hirata et al., 2006; Qin et al., 2009; Stenberg and Andersson, 1979; Tawancy, 1996) are all growth related and not induced by deformation.

In this work, the dominant experimental difficulty of brittle failure before significant plastic deformation at low temperatures is overcome by employing nanoindentation in conjunction with atomic force microscopy, electron backscatter diffraction (EBSD) and transmission electron microscopy (TEM). This combination allows the identification of active slip planes in this complex crystal and can therefore not only give a better indication of the mechanical anisotropy but also guides subsequent uniaxial tests to determine critical resolved shear stresses (CRSS) on particular slip planes (Korte-Kerzel, 2017; Korte and Clegg, 2012). Such tests were carried out using single crystalline micropillars, in which brittle failure is largely suppressed even in the most brittle and complex crystals (Korte-Kerzel, 2017; Korte-Kerzel et al., 2018), and selected orientations giving high resolved shear stresses on the slip planes identified by indentation and TEM.

## 2 Experimental Methods

### 2.1 Sample preparation

The sample composed of 45 wt.-% Fe and 55 wt.-% Mo was arc-melted from pure elements and subsequently re-melted and quenched 5 times to obtain a homogeneous composition and microstructure. An additional heat treatment of 1000 h at 800°C was performed to improve the chemical and microstructural homogeneity of the sample material. Small specimens with 1 mm thickness and a cross-section of approximately 10 x 5 mm were cut, mechanically ground and finally polished using 0.05 µm alumina suspension.

## 2.2 X-Ray powder diffraction and phase analysis

An X-ray diffractogram of a powderized sample was collected using a conventional diffractometer (Philips PW1820, Philips N.V.), operated in Bragg-Brentano geometry and equipped with a Cu tube, a secondary graphite (002) monochromator and a scintillation counter. A $\theta/2\theta$-scan was performed in the $2\theta$-range 10-110° in steps of 0.02°, counting 20 s/step. In addition to a 10-coefficient background polynomial and the height of the flat sample, lattice parameters for the observed phases $Fe_7Mo_6$, Fe, and Mo were refined. Line profile parameters could be refined with identical values for $Fe_7Mo_6$ and Fe, while for Mo an individual (sharper) peak shape was needed. No thermal displacement parameters were allowed to vary, while the fractional x- and z-coordinates of the $Fe^{2+}$ atom in $Fe_7Mo_6$ (Wyckoff position 18 h in the hexagonal setting of space group $R\overline{3}m$; original structural model by Lejaeghere et al. (Lejaeghere et al., 2011)) and the z-coordinates of the Mo-atoms in $Fe_7Mo_6$ were refined. The TOPAS suite of programs (Coelho, 2007) was used for the refinement, employing the Rietveld method (Rietveld, 1967).

## 2.3 Nanoindentation

Room temperature nanoindentation experiments were conducted using a Berkovich diamond indenter tip at 50, 100 and 200 mN peak load with constant loading and unloading rates of 1, 5, 10 and 20 mN/s and 5 s dwell time at maximum load (NanoTest Platform 3, MicroMaterials Ltd.). The load-displacement curves were analysed by the Oliver and Pharr method (Oliver and Pharr, 2004). The indentation modulus was determined using the Sneddon correction (Oliver and Pharr, 2004) with a Poisson ratio of 0.32 for the sample and 0.07 for the indenter, as well as a Young modulus of E = 1140 GPa for the diamond tip. Since there is no available data on the Poisson ratio of the µ-phase, this Poisson ratio was chosen according to first principles calculations by Huhn et al. (Huhn et al., 2014). Here we assume an average of their calculated Poisson ratios for the stoichiometric composition $A_6B_7$ for all calculated combinations, since no modulus match to the materials investigated in the present study is available. Taking a deviation of ±0.02 from 0.32 into account, the resulting error is in the range of <2%.

## 2.4 Surface microscopy

Optical light imaging (Leica DMR, Leica AG) and scanning electron microscopy (SEM) (Helios Nanolab 600i, FEI Inc.) were performed to characterise the morphology and topology of the indentation imprints. Surface topography scans of indents were performed using an atomic force microscope (AFM) in tapping mode (Dimension 3100, Bruker Corp.), followed by topography signal analysis utilizing the WSXM software (Horcas et al., 2007). All indents used for slip trace analysis were additionally examined by EBSD to facilitate a correlation of the surface morphology and crystal orientation.

## 2.5 Surface trace analysis

For the analysis of the activated slip systems, only indents carried out at 100 mN and 200 mN were used as hardly any lines were visible on the surface at a maximum load of 50 mN. Figure 2a shows a typical AFM image of the surface topography around a residual indent revealing that the surface traces around the indent contain different features. These features are classified into categories depending on their appearance. The first category contains straight, but shallow traces marked by the dashed lines in Figure 2a and the left inset of Figure 2a. These features are named "lines" throughout the text. Further, serrated and wavy traces were found which can be described by small straight segments as shown in the right inset (Figure 2a), these segments are called "edges" throughout the text. Other high and continuously curved traces, which inhibit accurate angle measurements, are excluded from this analysis.

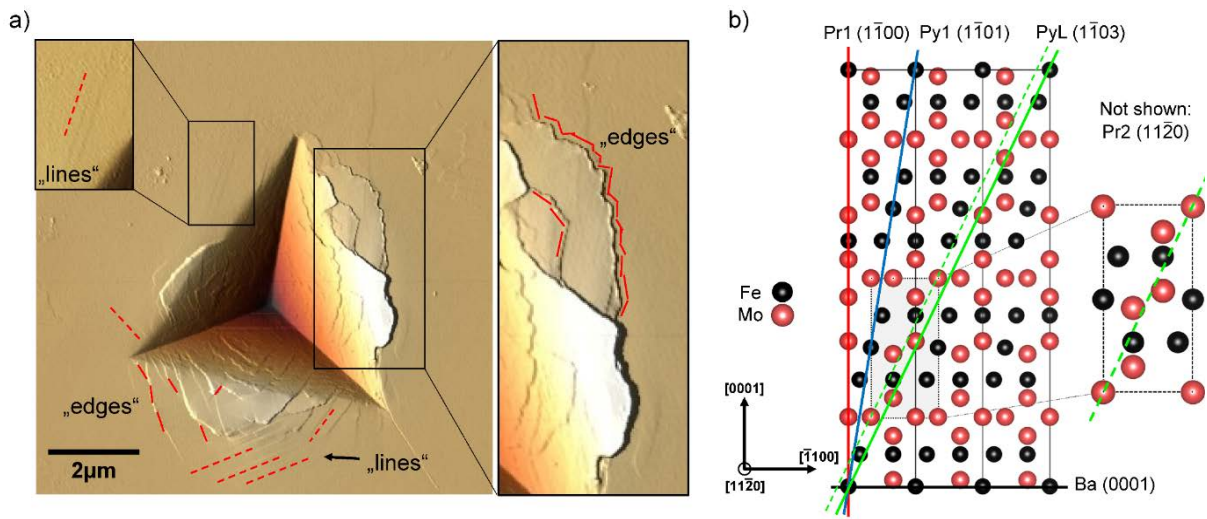

**Figure 2: (a) AFM topography scan (3D top view representation) of an indent showing different types of slip traces ranging from hardly visible straight lines ("lines"), to curved or serrated and wavy traces with in the majority much greater surface height. In some cases, the latter ones can be described by short, straight segments ("edges"); (b) Unit cell of the μ-phase indicating the slip planes considered for the surface trace analysis. The enlarged image of the Laves-phase subcell illustrates the pyramidal plane denoted as PyL, the first order pyramidal plane in the given Laves-type subcell.**

In order to correlate the observed surface traces of indentation imprints with possible glide (or crack) planes, a slip trace analysis similar to the one described by Nibur and Bahr for FCC crystals (Nibur and Bahr, 2003), Zambaldi et al. on α- and γ-titanium (Zambaldi and Raabe, 2010; Zambaldi et al., 2012) and Tromas et al. on MAX-phases (Tromas et al., 2011) was performed. Here, a statistical method for the plane-to-trace assignment was implemented to improve the interpretation of the results in the anisotropic μ phase, employing both scanning electron (SE) and atomic force imaging in combination with EBSD measurements. The planes considered for the following analysis are indicated in Figure 2b. Further, a schematic drawing of the surface trace analysis process is shown in *Figure 3*.

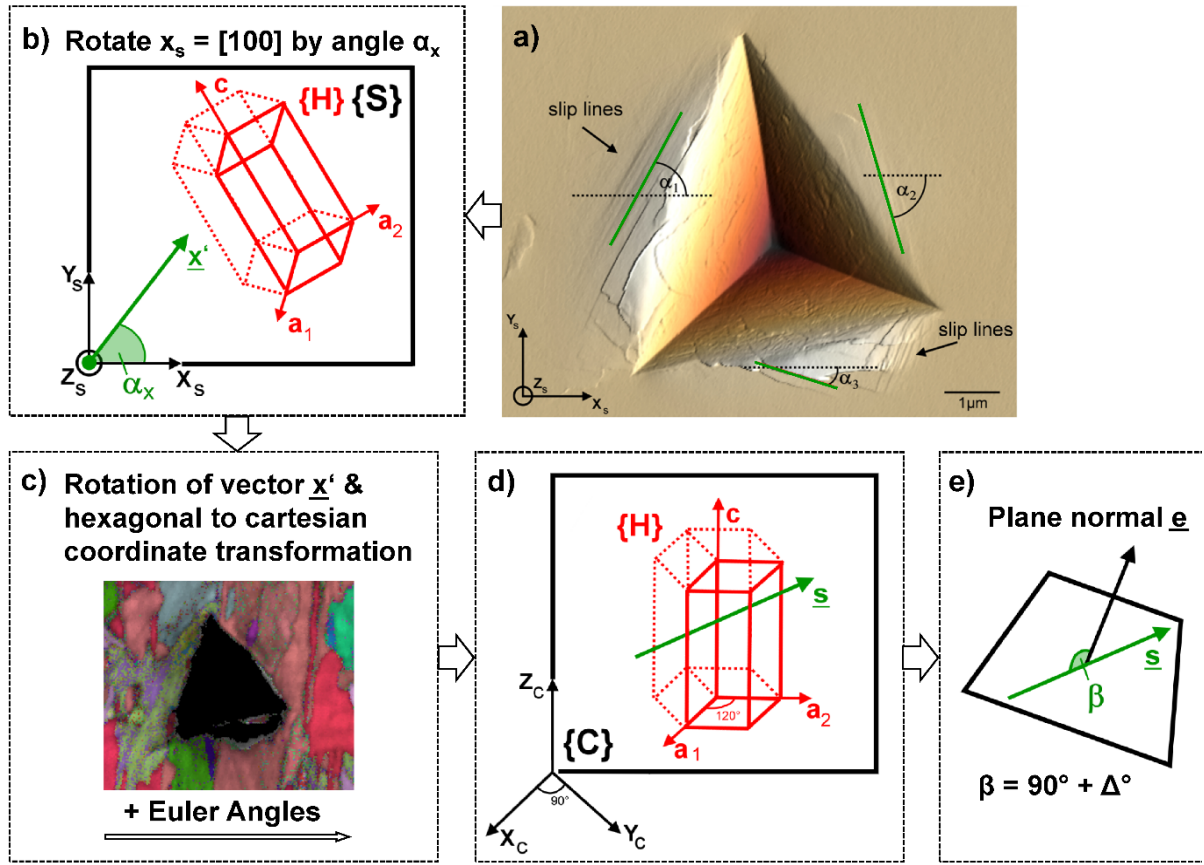

**Figure 3: Schematic representation of the surface trace analysis described: (a) angular measurement of traces and rotation of sample x-axis to obtain trace vector (b). Further rotation by Eulermatrix and hexagonal to cartesian transformation (c) give the vector in crystal coordinates (d). If the dot product of plane normal and vector equals 90° ± Δ the observed surface trace lies within plane (e).**

Visible surface traces and their angles $\alpha_x$ to the horizontal of the image (positive x-axis in samples coordinates) were measured from the SE and AFM images. Upward facing angles were denoted as positive and downward facing angles as negative. A vector $\vec{x'}$ (*Figure 3*a, b) following the surface trace on the sample surface, i.e. the x-y plane, is obtained by rotation of the horizontal sample coordinate axis [$\vec{x_s} = (1,0,0)$] by $\alpha_x$ around the sample normal, $\vec{z}$,

$$\vec{x'} = R_z(\alpha_x) * \vec{x_s} \tag{1}$$

with

$$R_z(\alpha_x) = \begin{bmatrix} \cos(\alpha_x) & -\sin(\alpha_x) & 0 \\ \sin(\alpha_x) & \cos(\alpha_x) & 0 \\ 0 & 0 & 1 \end{bmatrix} \tag{2}$$

The local crystal orientation in Euler angles $(\varphi_1, \phi, \varphi_2)$ at each surface trace was obtained by correlation of the AFM / SEM images with the corresponding EBSD maps (*Figure 3*c). Further rotation transforms the vector $\vec{x'}$ into crystal coordinates resulting in a vector $\vec{s}$

$$\vec{s} = R_{EBSD} * \vec{x'} \tag{3}$$

using the Euler-Bunge convention where $R_{EBSD}$ is the rotation matrix:

$$R_{EBSD} = \tag{4}$$

$$\begin{bmatrix} \cos\varphi_1 \cos\varphi_2 - \sin\varphi_1 \sin\varphi_2 \cos\phi & \sin\varphi_1 \cos\varphi_2 + \cos\varphi_1 \sin\varphi_2 \cos\phi & \sin\varphi_2 \sin\phi \\ -\cos\varphi_1 \sin\varphi_2 - \sin\varphi_1 \cos\varphi_2 \cos\phi & -\sin\varphi_1 \sin\varphi_2 + \cos\varphi_1 \cos\varphi_2 \cos\phi & \cos\varphi_2 \sin\phi \\ \sin\varphi_1 \sin\phi & -\cos\varphi_1 \sin\phi & \cos\phi \end{bmatrix}$$

The orientation of any given crystal plane can be described by its plane normal, $\vec{n}$, translated from Miller-Bravais to Miller indices using the relation (Nan et al., 2012)

$$hkil \rightarrow \left(2*h+k, 2*k+h, \left(\frac{3}{2}\right)*\left(\frac{a}{c}\right)^2 * l\right) \rightarrow \vec{n} \tag{5}$$

In Cartesian coordinates $l$ depends on the c/a ratio and is therefore further transformed to $\vec{e}$ (equation (6)), linking the hexagonal to the Cartesian coordinate system (Figure 3d) (equation (7)),

$$\vec{e} = \begin{bmatrix} x_E \\ y_E \\ z_E \end{bmatrix} = R_{cart} * \vec{n} \tag{6}$$

with $R_{cart}$ as the transformation matrix

$$R_{cart} = \begin{bmatrix} 1 & -0{,}5 & 0 \\ 0 & \frac{1}{2}*\sqrt{3} & 0 \\ 0 & 0 & \frac{c}{a} \end{bmatrix} \tag{7}$$

If a slip trace measured at the sample surface is formed by dislocation slip of a certain slip system, the slip trace must be in the slip plane. This condition is fulfilled when the trace vector, $\vec{s}$, is perpendicular to the plane normal $\vec{e}$, i.e. the scalar product of both is zero (*Figure 3*e). Due to the uncertainty in both indexing and measurement of the trace orientation, a deviation, $\Delta$, from the ideal solution of γ = 90° was included, where γ is the angle included between the vectors $\vec{e}$ and $\vec{s}$:

$$\gamma = \cos^{-1} \frac{\vec{e} * \vec{s}}{|\vec{e}| * |\vec{s}|} \tag{8}$$

The solution given in equation (8) is not unique since several planes within the unit cell may contain the slip trace. The analysis for a given slip trace includes all possible slip planes within the given range $\Delta$. If no slip plane with $|\gamma - 90°| \leq \Delta$ was found, the surface trace was denoted as a "zero solution".

## 2.6   Micropillar compression

Based on the results taken from nanoindentation and the surface trace analysis given below, suitably-oriented grains of the μ-phase were identified by EBSD. Micropillars with a width-to-height-ratio of 1:2-2.5 were then cut by successive annular milling steps using focussed ion beam to the final proportions and imaged before compression. The final diameter of the pillars was milled to be 2 μm at about 4-5 μm of pillar length, using a low beam current of 86 pA.

Early pillar tests demonstrated large displacement bursts in the load controlled experiments once slip initiated, therefore, in order to ensure that the strain was limited, the first, largest-diameter cut was made to 600 nm depth, equivalent to around 15% of total pillar strain. The subsequent cuts had a maximum outer diameter of 9 μm, i.e. less than the 10 μm flat punch used for testing. This way, micropillars with `typical' dimensions were manufactured, which ensured that the flat punch was caught on the outside edge, preventing the pillar from being deformed to large strains impeding the identification of slip traces by measuring their angle and orientation. For this, images of pillars taken after compression were aligned such that measurements of the slip trace angles perpendicular to the slip direction and tilted 45° towards the electron beam were enabled.

All compression experiments were conducted with a 10 μm diamond flat punch at a loading rate of 1 mN/s (iNano, Nanomechanics Inc.). The recorded load-displacement curves were corrected by the Sneddon correction according to Frick et al. (Frick et al., 2008). For stress determination, the diameter at the pillar top was used. Final CRSS values were obtained after calculation of the Schmid factor, $\mathbf{m}_S$, of the determined actual slip planes after analysis and using plausible Burgers vectors (given with each calculated Schmid factor below).

## 2.7   TEM

TEM investigations were carried out on lamellae of selected indents and micropillars prepared by focused ion beam milling (Helios Nanolab 600i, FEI Inc.) to obtain site specific electron transparent samples. The micropillars were cut in the centre in vertical cross-section approximately parallel to the slip direction and the position of the lamellae extracted from the plastic zone of indents is illustrated with each respective micrograph. TEM analysis was done at 200 kV (CM20, Phillips N.V.).

## 3   Results

### 3.1   Phase analysis

*Figure 1*b shows an electron micrograph of the material investigated revealing a dual-phase microstructure consisting predominately of a matrix phase and an intergranular second phase. EDX measurements of the matrix and the intergranular second phase confirmed that the matrix phase has the chemical composition of the $Fe_7Mo_6$ phase (μ) with a composition range of 52-57 wt.-% Mo throughout the sample, while the intergranular dark veins are iron-rich precipitates with up to 16 wt.-% Mo. This is consistent with the EBSD phase analysis (*Figure 1*b). Further, isolated areas of pure Mo were found (not shown). *Figure 1*c shows the powder diffraction pattern of the same material revealing peaks of the μ-phase, as well as of pure iron and molybdenum. No diffraction peaks of Laves-phase were measured in the sample. Rietveld

analysis with four-digit accuracy was performed to calculate the crystal parameters of the hexagonal unit cell to **a** = 4.7570(6) Å and **c** = 25.7530(4) Å giving a c/a-ratio of 5.4137 (*Figure 1*a). The bracketed numbers show the estimated standard error of the Rietveld method on the fourth digit. Utilizing the so refined unit cell data, EBSD analysis revealed that the μ-phase matrix was found to be polycrystalline with a wide variation of grain sizes and shapes, ranging from submicron sized grains to grains with a diameter of more than 10 μm.

## 3.2 Hardness and Indentation Modulus

Due to the dual phase microstructure of the sample, around 53% of the indents were found to be in the proximity of a secondary phase, i.e. within 3 μm to the long sides of the residual indent impression on the surface. These were excluded from the analysis, giving a total number of 474 indents in the $Fe_7Mo_6$ phase. The irregular grain size distribution resulted in the vast majority of indents being placed in multiple grains and the indentations discussed in the following are therefore not representative of individual crystal orientations.

The hardness and modulus measurements for all three series of indentation peak loads, i.e. indentations to 50, 100 and 200 mN, yielded similar results within the experimental scatter. The discussion of the extracted quantitative values is therefore focused on those indents performed with a maximum load of 100 mN. These indents present the best compromise between the visibility and availability of slip traces, achieved only at 100 and 200 mN maximum load, and the minimisation of the volume of the plastic zone in order to reduce the effect of secondary phases.

It should be noted that any influence of secondary phases lying underneath the surface cannot be characterized for a large number of indents, however, the obtained hardness values were used to evaluate the effect of a metallic component grain underneath the indent. .

Figure 4 displays the nanoindentation hardness and modulus values for the 100 mN peak load experiments. The red columns show the values for all indents before secondary phase correction (denoted as "all") over all loading rates and the black columns show the corresponding corrected values ("corr"), i.e. excluding indents within 3 μm of the secondary phase at the surface. A high scatter of the hardness before and after correction is evident and a final indentation hardness value of 11.7±1.0 GPa with a reduced standard deviation was obtained. The average indentation hardness across all three peak loads using only those indentations which are 3 μm or further away from secondary phases amounts to the same average value of 11.7±0.9 GPa. By considering the different final loading rates, given by the variation of peak load and loading rate, only a weak strain rate sensitivity could be detected, which also remained within the experimental scatter.

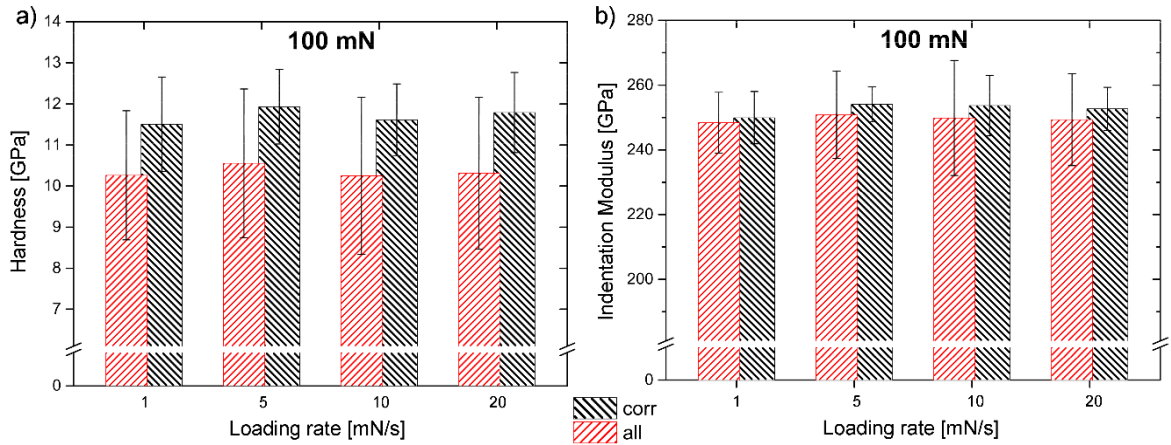

**Figure 4: Hardness (a) and indentation modulus (b) determined at a maximum load of 100 mN using loading rates of 1, 5, 10 and 20 mN/s, respectively. Both datasets are separated into the average of all indents measured (red colour, denoted as "all") and a reduced set of only those not in the proximity of any second phase visible at the surface (black colour, denoted as "corr"). Double standard deviations of the measurements shown as error bars.**

In contrast, the measured indentation modulus of the μ-phase (see Figure 4b for 100 mN max. load) did not show any significant variation before and after correction, exhibiting a slight increase of only 3 GPa after correction. This is not unexpected, as the measured hardness is directly dependent on the plastic zone around an indent and therefore highly influenced by the local (co-)deformation of a much softer secondary metallic phase, while the elastic zone is more widespread (Fischer-Cripps, 2006; Fischer-Cripps, 2007). The averaged value of the indentation modulus of the μ-phase was determined as 249±8 GPa and represents an average not only over the three peak loads investigated, which yielded no significant difference, but also the different crystal directions, for which anisotropy would be expected in measurements on single crystals.

### 3.3   Surface trace analysis

Figure 5 shows the results of the slip trace analysis for all investigated slip planes sorted by the defined surface trace categories ("lines", "edges"). The plane notations are indicated in Figure 2b and designated as "Ba" for basal, "Pr1" for prismatic 1st and "Pr2" for 2nd order prismatic (not shown). Further, pyramidal 1st order planes are denoted as "Py1" while PyL refers to pyramidal planes of 1st order within the Laves-phase subcells, as these were found to be potential slip planes of Laves-phases (Korte and Clegg, 2012; Takata et al., 2013). Lastly, "non" denotes "zero solutions", i.e. slip traces for which no matching plane could be found. Any further slip planes and associated directions, e.g. <c+a> on pyramidal planes of 2nd order were considered energetically highly unfavourable due to the prolonged unit cell in c-direction and therefore neglected in this analysis.

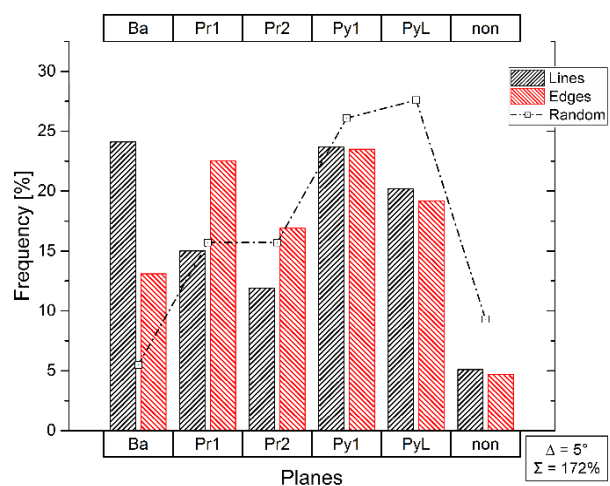

**Figure 5: Relative frequency of slip trace assignments categorized into "lines" (black) and "edges" (red) for an angle deviation Δ of 5°. The graph gives the total values of all slip trace assignments, annotated as "Ba" for basal, "Pr1" and "Pr2" for first and second order prismatic, "Py1" for pyramidal planes of first order. "PyL" represents the pyramidal planes of first order of the Laves-phase subcells, extrapolated to the full µ-phase unit cell and "non" for "zero solution", respectively. The dashed line gives the relative frequency of a random calculation of $10^6$ combinations, processed equally to the corresponding data.**

Angular deviations of Δ = 5° (Figure 5) and Δ = 3° were applied, however, Δ = 3° was found not to be appropriate due to experimental uncertainties, such as angular measurement and EBSD and AFM / SE image alignment, exceeding the deviation itself (see supplementary material for more detail on analysis with Δ = 3°). Within this study, a total amount of 272 clearly distinguishable traces was investigated, whereof 150 traces were classified as "lines" and 122 as "edges" (Figure 5).

In order to evaluate the significance of the relative frequencies of the indexed traces based on activation rather than the effect of symmetry within the unit cell, an additional theoretical statistical analysis was performed for $10^5$ crystal orientations and $10^2$ random slip trace angles. The resulting frequencies for each considered slip plane are given as a dashed line in Figure 5.

Due to the existence of ambiguous orientations, where a trace vector lies in more than one potential slip plane, and the included angular range, more than one possible solution is found per trace on average. By filtering the indexed planes to include only one assignment per plane family and using Δ = 5°, an overall indexing of Σ = 172% is reached.

Differences in frequency exist between the two types of surface traces and the crystallographic planes. Within the "lines" category, the basal fraction is about three times higher than in the "edges" category. For prismatic 1st order assignments the opposite trend is seen: a higher frequency within the "edges" category. When compared to the statistically random data, the basal "lines" and prismatic 1st order "edges" are significantly above the statistically random values. Although the assignment to pyramidal slip (Py1 and PyL) exhibit high fractions throughout this analysis, they are always lower, in case of PyL significantly lower, than the statistically calculated values.

The analysis was further confirmed for individual indents using TEM, see below.

### 3.4 TEM

As the majority of indents are placed in multiple grains, TEM lamellae have been taken from areas around indents loaded to 100 mN and 200 mN peak load where different resolved shear stresses (RSS) on the basal and prismatic planes common for hexagonal slip (Yoo, 1981) could be identified in individual grains encompassed by the indentation. In this context, the term RSS therefore refers only to the specific area/grain the lamella was taken from and is not applicable to the entire indentation into two or more grains. In the following, the results are presented in order of decreasing RSS on the basal planes and increasing RSS on the prismatic planes. Arrows in the figures denote the relative value of the resolved shear stress, the hardness value of the corresponding indent is also shown.

The lamella of the indent shown in Figure 6a has a basal plane inclination of about 35° to the indentation surface, and hence, the analysed grain possesses a high resolved shear stress on the basal planes. The surface topology is characterized by only few slip lines and high traces (split up, "edges") on one side of the indent. A crack is visible both in the SEM and the TEM image. The crack runs along a (presumably grown) twin boundary and intersects the surface near the corner of the indent. Dislocations are found predominantly on the basal planes, while no pronounced slip is observed on prismatic and pyramidal planes. The defects away from the indent are identified as twins and stacking faults and are assumed to be grown-in defects based on observation of an undeformed sample (see supplementary material).

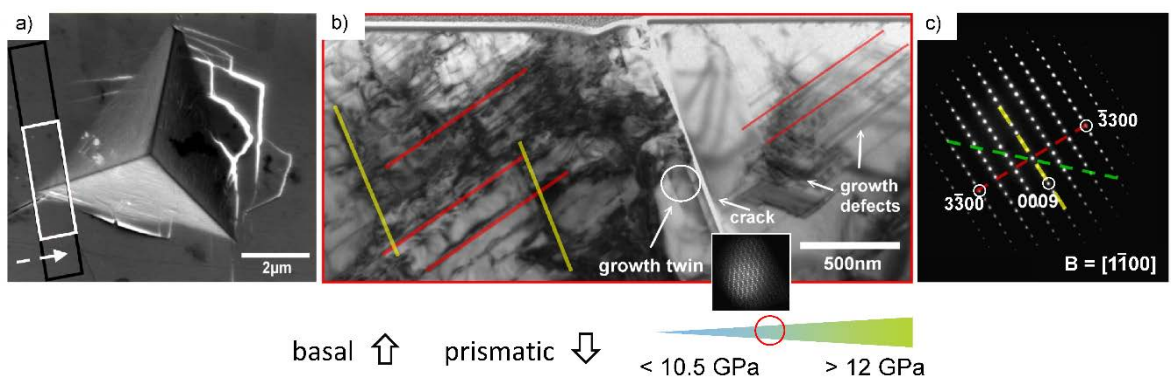

**Figure 6: SEM and TEM images revealing the surface topography and the corresponding microstructure underneath. Within the AFM / SEM images on the left (a), the alignment of the TEM-lamella and the field of view is given by the black and white rectangles, respectively. The arrow marks the direction of view. The BF-TEM image in the middle (b) shows the defect structure and slip planes in the plastic deformation zone of the indent. On the right, the corresponding diffraction pattern with incident beam and indicated plane orientations is given (c).**

Figure 7 shows an indent which is partly in a grain with a relatively high resolved shear stress on both, basal and prismatic, planes where a crystal rotation of about 10° is found underneath the indent as revealed from the corresponding selected area diffraction patterns (SADP) in Figure 7a. A strong bending of the slip planes towards the centre of the indent is evident. These bent planes are indexed as basal planes, while the slip planes which are inclined about 80° to the indenter surface are indexed as prismatic planes. The prismatic slip planes are seen as the higher and longer surface traces right and left of the indent, while the black arrows mark the traces extending from the basal planes in Figure 7a/b. The corresponding AFM analysis (Figure 7c) shows the surface profile of the steps formed by prismatic slip as illustrated by the red dashed lines.

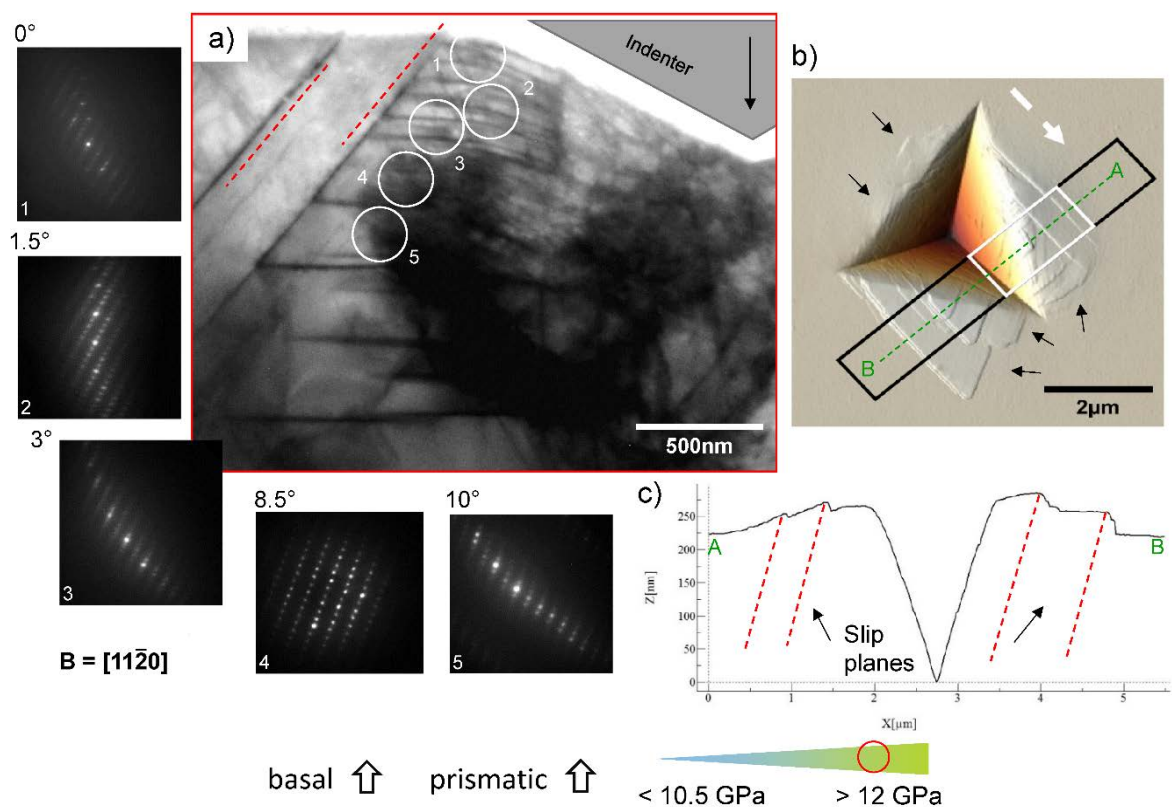

**Figure 7: (a) BF-TEM image showing the crystal rotation underneath an indent accompanied by plane bending of basal planes and the corresponding SADP. The alignment of the TEM-lamella and the field of view are given by the black and white rectangles in the AFM image in (b), the white arrow marks the direction of view. The black arrows point towards basal plane traces. The green dashed line specifies the location of the AFM height profile given in (c).**

Figure 8 shows an area from an indentation in which the resolved shear stress on the basal planes was low. The SEM image in Figure 8c reveals only minor surface traces formed on two sides of the indent. Similar to Figure 6a, a crack is observed emanating next to the bottom left corner of the indent. The TEM lamella was cut perpendicular to the pronounced traces on the right side of the indent. The active slip planes corresponding to the observed surface traces are mainly prismatic planes. Beside the highly deformed zone below the indent (right side of the TEM image), an area of finely spread prismatic slip planes with a lower defect concentration is observed. The SADP of this area (marked by the white circles) indicate the presence of stacking faults on the prismatic plane as evident from the pronounced streaking. These streaks are not present in the SADP of the areas without pronounced slip planes. Further, diffraction spots of a twin (presumably a grown-in twin) are observed. It should be mentioned that this region is connected to the plastic zone under the indent (Figure 8a), building a network of prismatic slip around the area with less defects.

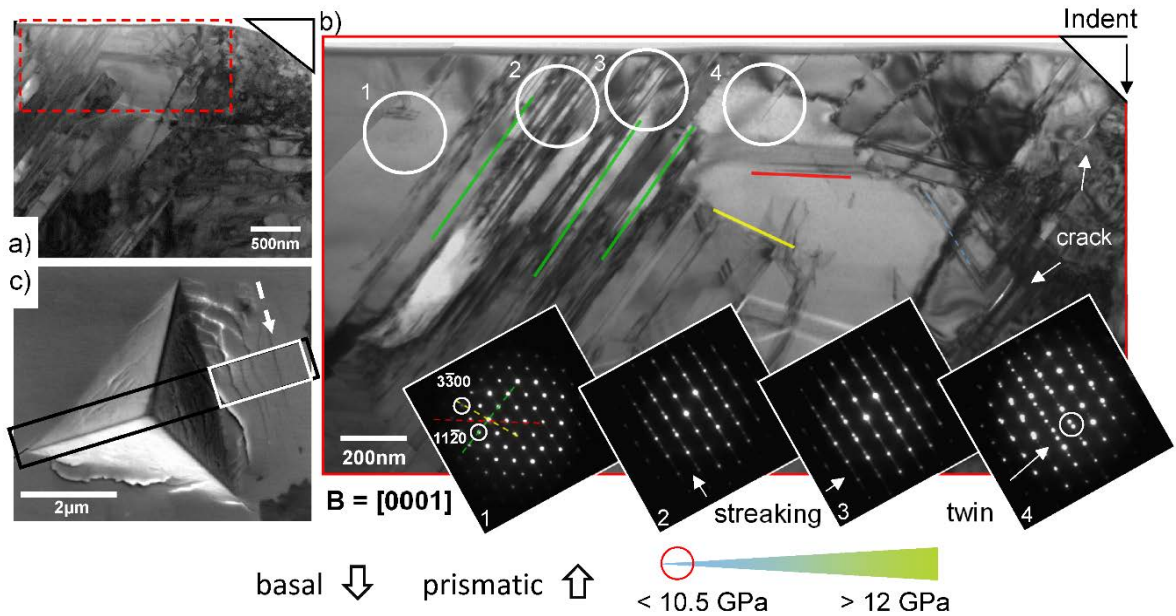

**Figure 8: (a) Low magnification overview TEM image showing the position of the more detailed BF-TEM image on the right (b) The white circles correspond to the SADP-patterns from left to right. (c) TEM lamella alignment as shown by the black rectangle in the SEM micrograph. The field of view is given by the smaller white rectangle on the right, the arrow points into the direction of view.**

Below the indent (right side of the micrograph), defect accumulation on first order prismatic planes is observed, these are intersected by defects on second order prismatic planes. As indicated by the arrow, the crystal is rotated leading to a bending of the slip planes in the plastic zone. Below the indent small cracks along first order prismatic planes can be seen.

The TEM lamella shown in Figure 9b was prepared from an indented area where the grains' basal plane is nearly parallel to the loading direction giving a low resolved shear stress on the basal planes. The lamella contains two high, "wavy" surface steps, i.e. "edges". The activated slip planes were indexed as first order prismatic planes. Large numbers of defects are concentrated on two prismatic planes extending to the observed surface traces. The corresponding angle of this prismatic plane is measured to be ~54° to the indentation surface, giving a high resolved shear stress on prismatic planes. Further, intersecting prismatic planes with high defect concentration nearly parallel to the surface or inclined towards the indent are observed. Within the area below the indent (see direction of the arrow in Figure 9b), the defect concentration is too high to identify individual planes.

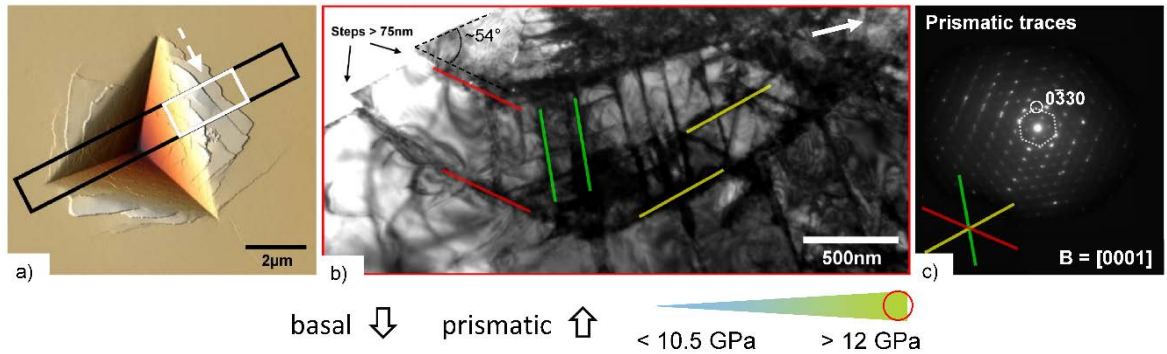

**Figure 9: AFM and TEM micrograph showing the surface topography and the corresponding microstructure underneath. Within the AFM image on the left (a), the alignment of the TEM-lamella is given by the black and the field of view by the white rectangle. The arrow marks the direction of view. The TEM image in the middle (b) exhibits the defect structure and slip planes in the plastic deformation zone of the indent. The corresponding diffraction pattern with incident beam and indicated plane orientation is shown on the right (c).**

During FIB sectioning for lamellae preparation, in some cases cracks have been observed to open (Figure 10, white arrows) emanating from high steps (inset Figure 10), characteristic for prismatic slip (Figure 9). Note that blunting of the steps given in the inset is due to prior FIB imaging. As can be seen, these cracks do not follow a straight crack path but rather have a wavy appearance. Angle measurements for an approximated straight crack path in the upper part, as shown in Figure 10, yield values of ~50° and ~59°. These angles roughly match the angles of the prismatic and pyramidal planes of the corresponding μ-phase unit cell to the horizontal sample surface.

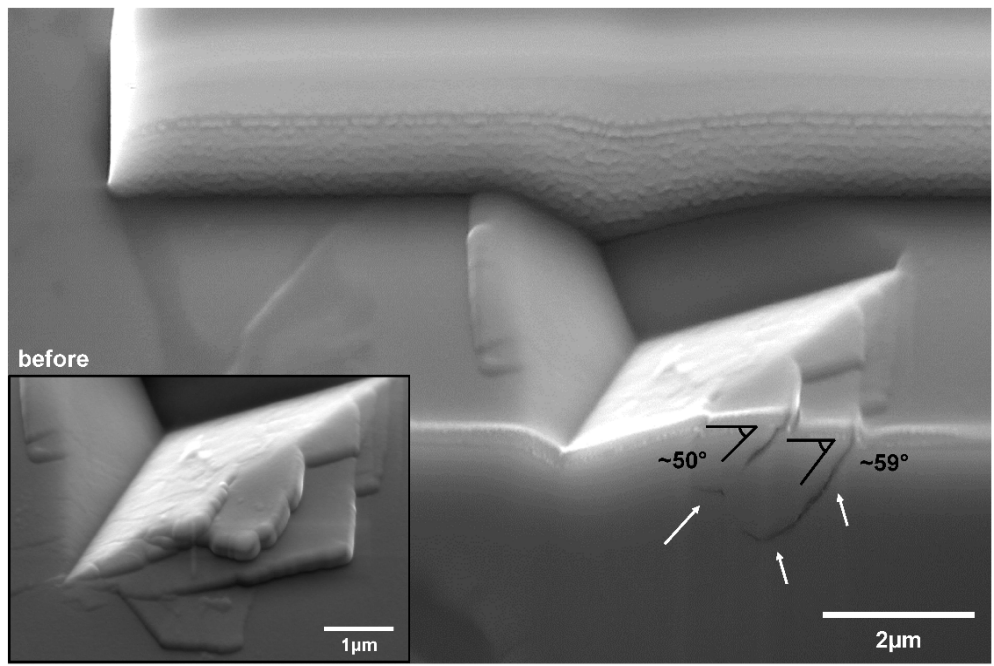

**Figure 10: FIB cross-section of high prismatic surface traces. Cracks running into the volume are highlighted by arrows. (NB: step blunting is due to prior FIB imaging)**

## 3.5 Compression of single crystalline micropillars

Based on the results of the surface trace and TEM analyses, as described similarly in (Korte et al., 2011; Soler et al., 2012; Wheeler et al., 2013), micropillar compression has been carried out on micropillars oriented for slip on basal or prismatic planes. For each of these slip planes, six pillars were compressed and analysed. The results are given in Figure 11 for a specific pillar oriented for slip on either plane, with each of the associated Schmid factors **m$_S$** of the respective pillar denoted in Figure 11a and b, respectively.

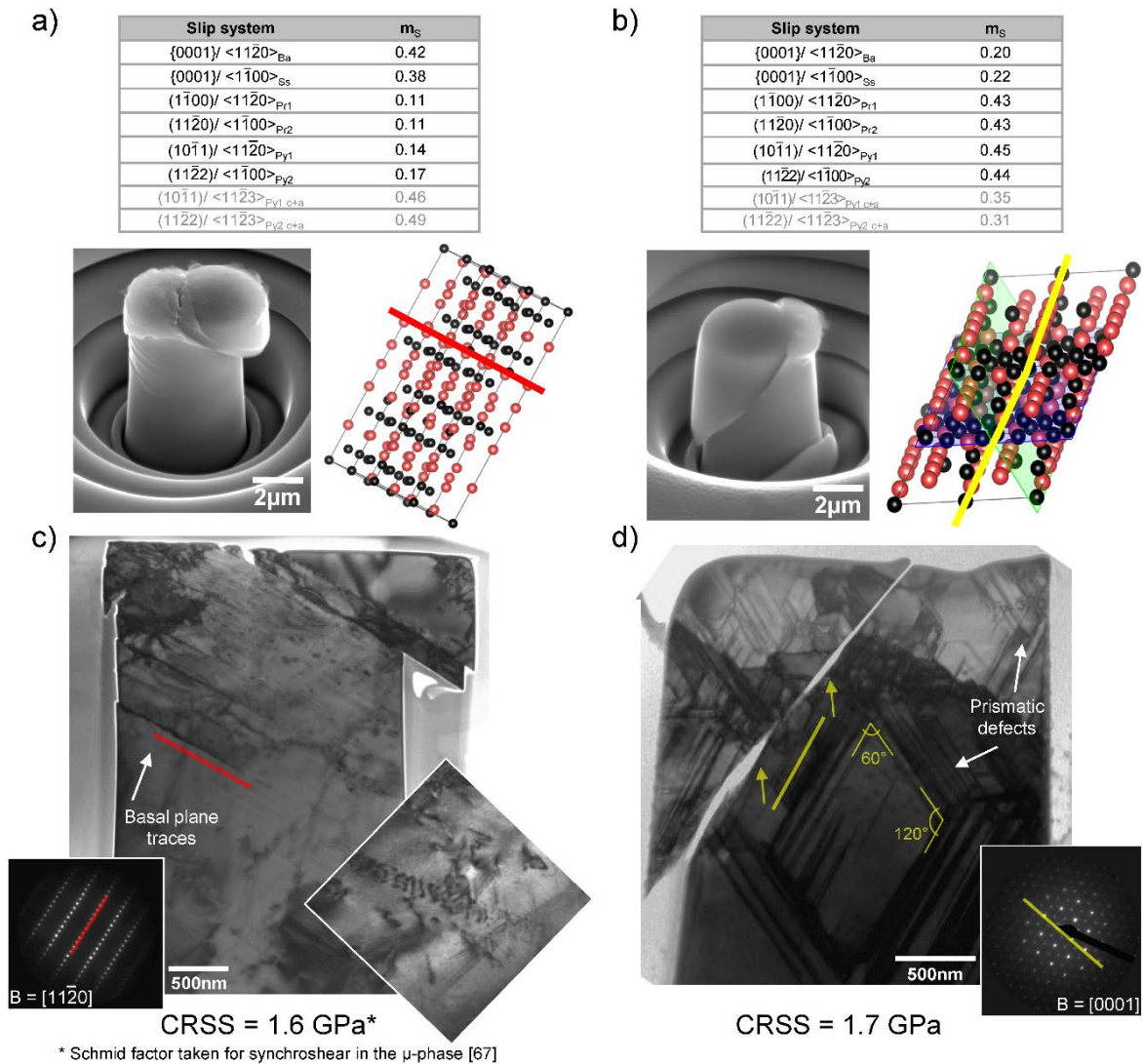

**Figure 11: Secondary electron micrographs of pillars oriented for (a) basal slip and (b) prismatic slip with the corresponding unit cell orientation indicating the observed slip plane and the corresponding Schmid factors, m$_S$, of the major possible slip systems in each pillar. TEM-micrographs of the (c) basal and (d) prismatic oriented deformed pillar reveal the corresponding defect structure. The slip planes are again indicated in the associated selected area diffraction patterns. The additional inset in (c) was taken after tilting away from the edge-on view of the slip plane in another pillar also oriented for basal slip.**

The slip directions for which Schmid factors were calculated were assumed either consistent to common hexagonal slip (Yoo, 1981) or, in consideration of the stacked-in Laves-phase subcell, with the Burgers vector direction of synchroshear (suffix Ss) through the Laves-phase triple layers, i.e. $\frac{1}{3}<1\bar{1}00>$ (Hazzledine and Pirouz, 1993; Hazzledine et al., 1992; Schröders, 2018). The Schmid factors of possible other slip systems are given also to indicate the relative stresses on each potential slip system.

In Figure 11a an electron micrograph showing a micropillar oriented for basal slip, the corresponding orientation of the unit cell and the relevant Schmid factors for the possible slip systems are given. Considering slip on only those planes activated in indentation, i.e. basal and prismatic, the highest Schmid factor for slip was found in **a**-direction followed by that for slip by synchroshear, also on the basal planes. Note that the resolved shear stress for pyramidal slip is even higher than for basal slip in this orientation. Consistent with the slip trace analysis, pyramidal slip was not expected or observed, neither "Py1" nor "PyL". For the micropillar orientation shown in Figure 11b, the basal slip systems exhibit the lowest Schmid factors, $\mathbf{m_S}$, while those of prismatic and pyramidal slip systems show similar and higher values. In each unit cell, the activated glide plane is indicated, i.e. the red basal plane trace in Figure 11a/ b and the yellow prismatic plane trace (Pr1) in Figure 11c/ d. The pillar oriented for basal slip (Figure 11) shows two clearly distinguishable slip traces emanating from the top part and smaller slip traces on the lower left side. The pillar oriented for prismatic slip (Figure 11b) only shows one clear slip trace crossing the entire pillar at a steeper angle, also emanating from the former top of the pillar.

In Figure 11c and d the corresponding TEM micrographs to Figure 11a and b, respectively, are given. The bright field image of the pillar oriented for basal slip (Figure 11c) reveals that deformation was indeed carried by dislocations (shown as inset) on basal planes, indicated by the red line in the micrograph and the relevant SADP on the left. The defect density decreases away from the top of the pillar, typical for the slightly tapered geometry and consistent with a decreasing stress level. At the top left of the pillar in Figure 11c, the compression punch further deformed the exposed material edge after slip took place on the basal planes, resulting in additional deformation on non-basal planes in this highly stressed region experiencing non-uniaxial stresses after the initial slip event. The determined CRSS at the onset of deformation for this pillar is 1.6 GPa, assuming that synchroshear is operating on basal planes, in particular in the Laves-phase triple layers, as shown by Takata et al. (Takata et al., 2013) for a C14-Laves-phase or in particular for the Laves-phase-layers in the μ-phase by Schröders (Schröders, 2018).

A contrasting picture of deformation is revealed by the TEM micrograph of the pillar nominally aligned for prismatic slip in Figure 11d. Here, no clear slip traces can be identified and, although the compressed pillar in Figure 11b appears similar to that in Figure 11a in terms of the appearance of the slip trace at the surface, TEM reveals that a crack is running through the pillar which follows in part the prismatic traces indicated by the yellow arrows and solid yellow lines, respectively. Numerous defects of prismatic origin are visible throughout the pillar, with angles of 60° and 120° seen between the defects. However, no clear indication of mobile dislocations can be obtained in this pillar. For this specific pillar, the critical stress for the onset of plasticity was determined to be 1.7 GPa, taking the Schmid factor, $\mathbf{m_S}$, for the prismatic plane and deformation in the **a**-direction into account for calculation.

The determined engineering stress-strain curves of all compressed pillars are given in Figure 12a for basal oriented pillars and correspondingly for prismatic oriented pillars in Figure 12b. While the stress ranges in which both types of pillar start to deviate from elastic loading are comparatively similar, small differences in the slip behaviour are evident. Some of the pillars oriented for basal slip show much more stable flow than any of those oriented for prismatic slip. In addition, a significant hardening is observed after small initial displacement bursts in the pillars oriented for basal slip, while no or only a small rise in stress level is seen in the pillars oriented for prismatic slip, consistent with the large displacement burst observed in all of these pillars. The stable yielding with increasing stress level is most pronounced for the corresponding curve of the pillar shown in Figure 11a, marked by the red, enlarged curve. In both graphs, one curve notably stands out showing significantly higher stresses. These two pillars slipped (or broke, for the prismatic pillar) away from the top. Therefore the engineering stress, as typically calculated from the top diameter, is not representative of the stress experienced locally at the onset of deformation, the real stress in the failed region, estimated using the diameter at the top end of the slip trace as observed in the SEM, is much lower. Nevertheless, for prismatic slip, this value remains higher than the majority of measured values.

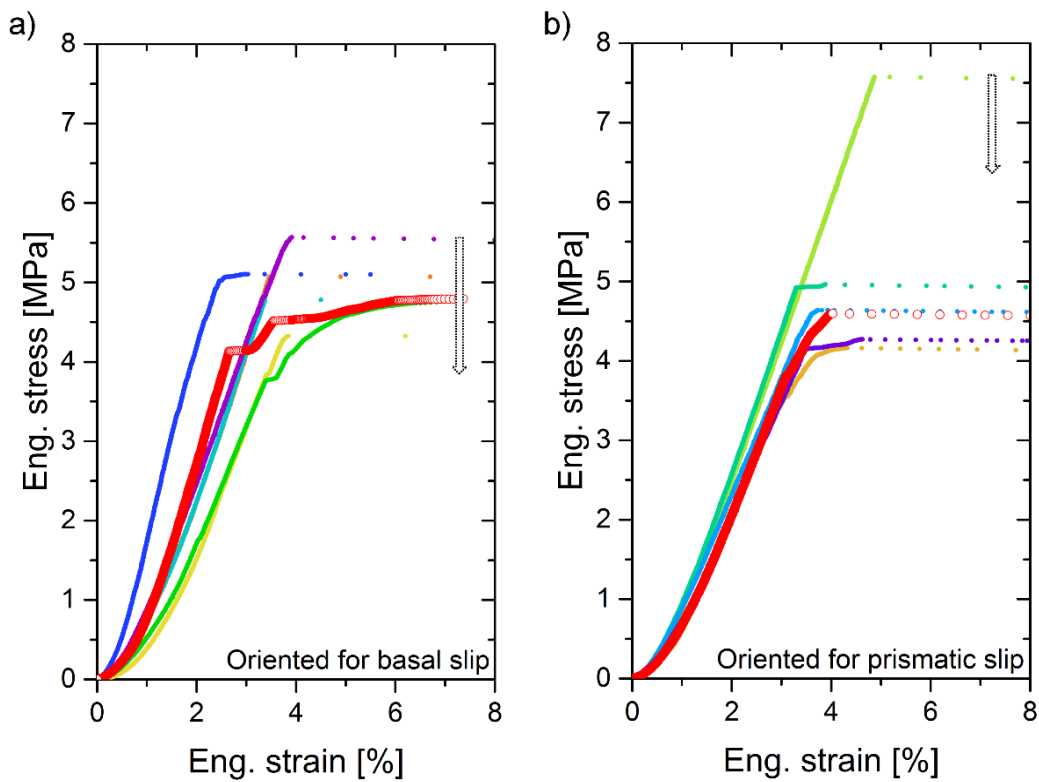

**Figure 12: Stress strain curves obtained from micropillar compression of pillars oriented for (a) basal and (b) prismatic slip. The red and slightly enlarged curves are the corresponding curves to the pillars shown in Figure 11a and b. One pillar in each set exhibits a higher stress level. SEM analysis revealed that in these cases slip was not initiated at the top of the pillar but further down. The arrows show the approximate value to which the stress might be corrected if the pillar diameter at the slip initiation site is considered for the calculation of the stress (instead of the top diameter).**

In Table 1, the average values of the CRSS calculated from the stress-strain curves are summarised. In the case of prismatic-oriented pillars, it cannot be decided whether this is representative of a fracture stress following a planar defect on prismatic planes or the onset of deformation by dislocation motion followed by decohesion of the slip plane and escape of the stored dislocations into the crack where three surfaces are in close proximity. (Note that both kinds of deformation, cracking along a pre-existing planar defect and decohesion along a plane with high accumulated defect density were observed in the TEM analysis of plastic zones underneath indentations). Depending on which slip direction is considered for basal slip in the pillars oriented for basal slip, i.e. the operating slip mechanism being either normal dislocation glide or synchroshear (Schröders, 2018; Yoo, 1981) , the Schmid factor $m_S$ is in the range of 0.41-0.45. Considered separately in terms of critical failure/ fracture ($\tau_F$) and the onset of plasticity ($\tau_Y$), this leads to CRSS values for basal slip (Ba) of $\tau_F = 2.05\pm0.24$ GPa  and $\tau_Y = 1.92\pm0.22$ GPa respectively. The corresponding CRSS values for synchroshear operation (Ss) are given as $\tau_F = 1.91\pm0.30$ GPa and $\tau_Y = 1.79\pm0.31$ GPa . The larger standard deviation in the latter case is the result of a greater scatter in Schmid factor for this slip direction. In comparison, the prismatic oriented pillars show higher CRSS for fracture of $\tau_F = 2.24\pm0.57$ GPa as well as at the onset of deformation (by whichever mechanism) of $\tau_Y = 2.11\pm0.66$ GPa, accompanied by a higher deviation. The average Schmid factor $m_S$ of $0.45\pm0.02$ for prismatic slip is identical to the one determined for normal basal slip (Ba) in **a**-direction in the basal oriented pillars.

Table 1: Average values of CRSS for fracture ($\tau_F$) and onset of plasticity ($\tau_Y$) with the associated Schmid factors for the slip systems found by nanoindentation and TEM analysis. Note that from the analysis no differentiation between the slip mechanisms considered for basal slip (Ba and Ss) could be made.

| | Slip systems | | |
|---|---|---|---|
| | of pillars oriented for basal slip | | of pillars oriented for prismatic slip |
| Average of | $\{0001\}/ <11\bar{2}0>_{Ba}$ | $\{0001\}/ <1\bar{1}00>_{Ss}$ | $(1\bar{1}00)/ <11\bar{2}0>_{Pr1}$ |
| CRSS: $\tau_F$ | 2.05±0.24 GPa | 1.91±0.30 GPa | 2.24±0.57 GPa |
| CRSS: $\tau_Y$ | 1.92±0.22 GPa | 1.79±0.31 GPa | 2.11±0.66 GPa |
| Schmid factor $m_S$ | 0.45±0.02 | 0.42±0.04 | 0.45±0.02 |

## 4  Discussion

### 4.1  Hardness and indentation modulus

The results obtained here can be readily compared with the literature. With a nominal stoichiometric composition $A_7B_6$, the µ-phase occurs in a wide range of chemical compositions incorporating the alloying elements within superalloys (Cheng et al., 2012; Krishna et al., 2013; Rae and Reed, 2001; Rehman et al., 2015; Yang et al., 2006) but also in a number of binary systems (Carvalho et al., 2000; Huhn et al., 2014; Wagner et al., 1995).

Rehman et al. (Rehman et al., 2015) have examined chromium, cobalt and rhenium rich μ-precipitates by AFM indentation in a creep deformed Ni-based superalloy. The obtained hardness values are slightly higher than the ones obtained in this study, giving 14.5±2 GPa, as well as a reduced modulus of 220±20 GPa, which corresponds to an indentation modulus of the order of that measured here, assuming the same Poisson ratio of 0.32. A deviation would in any case be expected simply as a result of the different elements incorporated (see also below) with the reported values likely representing a lower bound, i.e. the precipitate might possess a higher stiffness than measured, due to the effect of the elastic zone extending into the superalloy, which gives a slightly lower modulus in most indentation directions (Takagi et al., 2004). When considering the small maximum indentation load and depth of approximately 400 μN and 100 nm, respectively, the higher hardness at a similar modulus to that measured here may be due to a size effect on the hardness towards very small depths (Durst et al., 2008; Fleck et al., 1994; Nix and Gao, 1998; Pharr et al., 2010) or the result of the different composition of the phases tested. Note that the effect of the surrounding volume with a smaller hardness is not as pronounced in case of the hardness as the elastic field of the indentation is much larger than the plastic zone.

Instrumented indentation experiments at slightly higher loads of up to 1 mN have been carried out on the μ-phase's archetype $Fe_7W_6$ by Matějíček et al. (Matějíček et al., 2015), yielding an indentation modulus of 356±16 GPa and a hardness of 22.3±1.3 GPa. Their (Matějíček et al., 2015) results are in good agreement with ab-initio calculations by Ren et al. (Ren et al., 2014) giving a modulus of around 367 GPa for the iron-tungsten phase. Huhn et al. (Huhn et al., 2014) showed by first principles studies on early-late transition metal alloys and experiments on Ni-Ta alloys that the stiffness of $A_7B_6$ phases is slightly higher than that of the pure elements where the elastic constants or moduli considered are similar. Where the difference is large, the stiffness of the $A_7B_6$ phase falls in between those of the pure phases. This finding is consistent with the modulus of the $Fe_7W_6$ phase, $E_{Fe_7W_6}$=356 GPa (Matějíček et al., 2015), with the moduli of the pure elements of $E_{Fe}$=211 GPa and $E_W$=411 GPa. Similarly, the indentation modulus of 249 GPa found here for the $Fe_7Mo_6$ μ-phase falls in between that of iron and molybdenum ($E_{Mo}$=325 GPa (Gale and Totemeier, 2003)) and the modulus of the μ-phase measured by Rehman et al. (Rehman et al., 2015) with W, Cr, Co and Re ($E_{Cr}$=279 GPa, $E_{Co}$=211 GPa, $E_{Re}$=466 GPa (Gale and Totemeier, 2003)) as the major elements falls short of their average Young modulus of 340 GPa.

Here, most indents were placed into several grains, yielding a good averaging over the elastic properties but maintaining relevance to the orientation dependent plastic properties. Consequently, a correlation of the dominant surface trace appearance and hardness can still be identified and yields the following trend:

Pronounced slip traces on more than one side of the indent (Figure 9a) correspond to higher hardness. A lower hardness, by approx. 1 GPa, is found where less and lower surface traces are formed (Figure 6a). In particular, indents showing intersecting slip planes (Figure 7a) or deformation in which defects are concentrated on only a few distinct planes heavily occupied with dislocations in TEM (Figure 9b) and visible as high steps in AFM (Figure 9a) exhibit slightly higher hardness values. From correlation of orientation dependent slip morphology on the surface and in cross-section with the measured hardness, we may therefore deduce that hardening occurs at slip plane intersections and that the appearance of slip confined to individual planes is related to a hardening effect as well. However, due to the complex stress state underneath the indenter and the small grain size of the sample relative to the indentation size, the plastic as well as elastic anisotropy of the hexagonal unit cell cannot be accounted for in a quantitative manner based on indentation and microscopy data alone. The plastic anisotropy may, however, be assessed more quantiatively based on the uniaxial microcompression tests (section 4.3). As the crystal orientations for these tests were chosen following the identification of the dominant active slip planes, these will be discussed in the following section.

## 4.2 Active slip planes

The active slip planes in the $Fe_7Mo_6$ μ-phase were studied by surface trace analysis as a statistical approach for a large number of indents and TEM for selected indents for validation and further analysis of the deformation mechanisms.

For an approximately random distribution of grain orientations covered by the analysed indentations (see supplementary material), deformation was found to occur predominately along basal and prismatic planes. The high (absolute) frequency of pyramidal planes indexed during the analysis could be explained by comparison with a statistically random distribution of activated slip planes, which highlighted that this is a geometric effect, giving a (relative) advantage to the pyramidal planes owing to the crystal symmetry.

TEM confirmed the results of the slip trace analysis on individual indents with dislocations found mainly on basal and prismatic planes. In grains with a high resolved shear stress on the basal plane, basal slip was predominantly activated. Of course, due to the complex stress field below a Berkovich indenter and the associated activation of different slip planes, it is difficult to make exact statements about the CRSS. However, basal slip appeared to be suppressed in grains where the basal plane was aligned parallel or perpendicular to the indentation direction. This suggests anisotropy with a slightly lower CRSS for basal slip.

No strain-induced defects on pyramidal planes were found using TEM, consistent with the slip trace analysis, where no significant activation of slip on pyramidal planes above the statistically random distribution was observed. This may indicate a high CRSS and consequently infrequent or no activation of pyramidal slip. However, the high defect density in the plastic zone underneath the indent prevented the determination of single defect orientations. Pyramidal stacking faults were observed in both, deformed and undeformed samples and have also been reported before to form in the μ-phase (Carvalho and De Hosson, 2001; Carvalho et al., 2000).

The TEM analysis further revealed a correlation between the density of slip bands and the density of grown-in stacking faults on basal and prismatic planes. In grains with a high density of grown-in stacking faults on the activated glide plane, the number density of slip bands is high and the number of dislocations per slip band is comparably low. In contrast, in grains which exhibit only few grown-in stacking faults on the activated slip plane, deformation is highly concentrated on few slip bands which exhibit a high dislocation density and high local strain (as evident from SADP analysis in Figure 7a). This indicates that these grown-in stacking faults might either act as nucleation sources for mobile dislocations or that slip along these stacking faults is structurally and energetically easier than in the perfect lattice. The exact mechanism, however, requires further investigation.

Previous studies have focused on the characterization of growth-defects in the μ-phase (Carvalho and De Hosson, 2001; Carvalho et al., 2000; Hiraga et al., 1983; Hirata et al., 2006; Qin et al., 2009; Stenberg and Andersson, 1979; Tawancy, 1996). However, the focus of the present study is the investigation of the deformation mechanisms. While growth-defects may influence the nucleation, mobility and motion of mobile defects during plastic deformation and consequently the hardening behaviour of the material, mobile defects govern the plastic deformation within the crystal. To distinguish the mobile defects from growth-defects, TEM analysis of undeformed material was performed (see supplementary material). This analysis shows parallel arrays of fine stacking faults and networks of grown-in stacking faults on different planes. Furthermore, growth-twins were found in the undeformed material. This is in good agreement with the results of Carvalho et al. (Carvalho and De Hosson, 2001; Carvalho et al., 2000), Stenberg et al. (Stenberg and Andersson, 1979), Qin et al.(Qin et al., 2009), Tawancy (Tawancy, 1996), Hirata et al. (Hirata et al., 2006) and Hiraga et al. (Hiraga et al., 1983), who also observed grown-in twins and stacking faults. In contrast to these growth-defects, the deformation induced defects observed in our study are concentrated in defined slip bands/ deformation bands and induce significant lattice rotations.

Bringing together the findings from hardness measurements, trace analysis and TEM, the following correlations may be identified: the highest hardness is observed where the resolved shear stress on the basal planes is low and where high steps indexed as prismatic planes are observed at the surface. This is consistent with the CRSS on the prismatic planes being higher than on the basal planes. In addition, the TEM observations have shown that high surface steps are formed where dislocations concentrate on prismatic slip planes intersecting with another slip plane in the volume. Dislocation motion is impeded by such intersections as seen in Figure 7a, also consistent with these structures corresponding to the indents of highest hardness and therefore significant hardening at intersections of slip planes with other slip planes or planar growth-defects.

On the other hand, the lowest hardness values were found in those indents where fine lines appeared on the surface which were indexed as basal planes. This was the case where the grain was oriented for basal slip, i.e. with a high resolved shear stress on the basal planes.

Due to deformation of several grains of different orientations per indent and pre-existing defect structures in the volume, i.e. wide or densely spread growth-defects on prismatic or basal planes, an intermediate hardness is observed in most cases. Such intermediate hardness values would therefore correspond both to a mixture of the above mechanisms and glide on prismatic planes without intersection with other defects.

Fracture was observed in some indentations and the cracks were found to, rather untypically, not emanate exactly from corners of the indents. This may be explained by preferential crack paths available close to the corner. Here, two possible paths have been observed: fracture along growth-defects (Figure 6b) and fracture along planes with high dislocation density indexed as prismatic planes, such as those seen in Figure 9b. While in the first case, a straight crack path would be expected and is in fact observed (Figure 6b), in the latter case deviations from a straight path are possible due to inhomogeneous defect concentrations, plane intersections and lattice rotation (Figure 8b). Such a crack is shown in Figure 10. Although crack nucleation from either the surface or intersecting slip planes in the volume could not be differentiated here, crack nucleation from intersecting slip planes has been reported previously, for example in MgO (Cottrell, 1961). Growth defects and planes occupied by a large number of dislocations are therefore implicated as sites of a lower cohesive strength. Their role in deformation of the μ-phase beyond its elastic limit is discussed further below, in the context of uniaxial deformation in micropillars.

## 4.3   Compression of single crystalline micropillars

In indentation, lower hardness for indents with mainly more shallow basal traces and higher hardness for indents with higher and more serrated steps of prismatic origin has been measured, while also fracture along prismatic planes has been observed. This raises the following major questions: First, can CRSS values be determined and is there indeed a measurable difference showing anisotropy with a lower CRSS on the basal planes? How do plasticity and fracture compete in the deformation of small volumes of µ-phase (precipitates or micropillars)?

For the case of slip on basal planes, dislocation mediated plasticity was shown for nanoindents and was also observed in micropillar experiments where dislocations have been identified on the basal planes and are thought to be strain-induced, as shown in the inset of Figure 11a. We therefore conclude that the critical stress value measured at the onset of plastic deformation does indeed correspond to the CRSS for dislocation glide on basal planes. However, a different CRSS is derived depending on which basal slip system is considered, conventional hexagonal slip on $\{0001\}/\frac{1}{3}<11\bar{2}0>_{Ba}$ (Yoo, 1981) or synchroshear $\{0001\}/\frac{1}{3}<1\bar{1}00>_{Ss}$ (Chisholm et al., 2005; Hazzledine and Pirouz, 1993; Hazzledine et al., 1992; Schröders, 2018). The operation of synchroshear needs to be considered in the µ-phase as individual constituents of its large unit cell may determine how flow occurs (Howie et al., 2017), namely the $MgCu_2$-type Laves-phase or the $Zr_4Al_3$ layers. Besides the crystallographic structure, nothing has been reported for the $Zr_4Al_3$ phase (Wilson et al., 1960), while several studies have been conducted for Laves-phases (Korte and Clegg, 2012; Paufler, 2011; Takata et al., 2013), also in terms of their deformation including micropillar compression. At room temperature, the synchroshear mechanism was proposed to operate in these (Chisholm et al., 2005; Hazzledine and Pirouz, 1993; Hazzledine et al., 1992), shearing the triple-layers of Laves-phases on the basal plane by synchronous movement of two co-planar partials. Takata et al. (Takata et al., 2013) identified synchroshear as the operative mechanism in their study on C14 $(Fe, Ni)_2Nb$ micropillars by assuming the direction of slip in the micropillar to be $<1\bar{1}00>$, consistent with the synchroshear vectors (Chisholm et al., 2005; Hazzledine and Pirouz, 1993; Hazzledine et al., 1992; Schröders, 2018). On the other hand, Korte et al. (Korte and Clegg, 2012) showed a slip direction consistent with conventional basal slip in micropillars of the cubic $NbCo_2$-Laves-phase, although a more lateral component, as expected for synchroshear, may not be decisively excluded from their top-view micrographs of deformed micropillars.

Unfortunately, for the micropillar compression experiments conducted here, the direction of slip could also not be unambiguously determined, as the angle difference between both slip directions is only 30° and lateral shifts of the slipped parts impede accurate evaluation. Further, inaccuracies from pillar rotation and unit cell orientation by EBSD need to be considered. However, by HR-TEM investigations of the same µ-phase studied here, evidence of synchroshear in the Laves-phase subcells was identified (Schröders, 2018). Although the operative slip mode of deformation on basal planes can therefore not be clearly distinguished here, plastic deformation by dislocation slip on basal planes has been shown by nanoindentation and micropillar compression in this study. Therefore, the average CRSS values for the onset of plasticity by dislocation slip on basal planes $\tau_Y$ can be given as 1.92±0.22 GPa, assuming basal slip (Ba), and 1.79±0.31 GPa (Ss), assuming slip to occur by synchroshear, (Table 1).

For the case of deformation on prismatic planes, slightly higher average critical stresses for the onset of plasticity, $\tau_Y$ = 2.11±0.66 GPa, and failure, $\tau_F$ = 2.27±0.59 GPa, were determined. Although fracture of pillars was clearly shown by TEM in Figure 11b, the majority of the stress-strain curves in Figure 12b show some stable plastic flow before the unstable deformation burst occurs. We therefore deduce that the pillars deform first by dislocation motion on prismatic planes followed by decohesion of the slip plane. This yields a consistent picture with the analysis of indentations by TEM above and implies that dislocation slip on prismatic planes can be activated at a similar or slightly elevated stress level compared with basal slip. However, where defects accumulate and no or only a small confinement exists, the prismatic slip planes become preferential cracking sites which fail at a critical stress of the order of the stress for dislocation motion.

Combining the observations made by nanoindentation, TEM and micropillar compression, we propose the following mechanisms underlying plastic deformation on prismatic planes: While loading along prismatic planes, dislocations move, presumably where defects are already present before, and accumulate on these planes. This is consistent with the observations in the lamella shown in Figure 8 where existing defects lead to localized dislocation slip. Under confining pressure, further pronounced slip on prismatic planes and defect accumulation (Figure 9b) proceeds with increasing stress, resulting in mutual obstruction of prismatic slip planes (Figure 9b). If the stress level exceeds the fracture stress locally, possibly due to locally very high defect concentrations introducing local stresses while lowering the cohesive strength, critical failure by fracture results. This is consistent with dislocation slip and defect concentration shown in the TEM micrographs of Figure 8b, where high defect concentrations and intersections result in cracks along prismatic planes underneath the indent, as well as the serrated crack paths along free surfaces, e.g. Figure 10. Further, yielding of prismatic oriented micropillars is evident in the stress-strain curves given in Figure 12b resulting in fracture at small strains compared to basal oriented pillars. This strongly supports the assumption that the onset of plasticity is governed by dislocation slip followed by dislocation accumulation on prismatic planes and ultimately fracture at a critical stress and defect density level.

In addition, the results obtained by nanoindentation and microcompression both indicate that defects and their density have a strong influence on the deformation behaviour of the μ-phase. This is also apparent in the stress-strain curves (Figure 12) given for both types of pillar. One curve each shows significantly higher stress values. The corresponding micropillars accordingly show slip initiation in the middle or lower part of the pillar. As indicated by the arrows, re-calculation of the stress values with the apparent pillar diameters leads to the same or slightly lower stress level for the basal oriented pillar, while the prismatic pillar still exhibits a higher stress value. This is in agreement with a strong interaction of slip planes and growth defects of prismatic and/or basal character. These may act in opposite direction depending on their relative orientation: Where the growth defect coincides with a potential slip plane experiencing a high resolved shear stress, the pre-existing defects appears to act as a preferential deformation site, presumably as the critical stress for dislocation nucleation and/or motion is reduced. Conversely, where the resolved shear stress is high on another plane which is intersected by growth defects, these obstruct dislocation motion and thus lead to hardening.

The obtained results may be used as a starting point to improve previous crystal plasticity simulations either of pure single or polycrystals (Roters et al., 2010; Segurado et al., 2012; Zecevic et al., 2017) or for the case mentioned here in modelling of co-deformation in microstructures containing μ-phase precipitates within a Ni-base superalloy (le Graverend et al., 2011). With the provided knowledge of primary deformation mechanisms, enhancement of existing deformation modelling efforts by le Graverend et al. (le Graverend et al., 2011) would be possible as, so far, these could only be based on grown-in defects (Carvalho et al., 2000; Qin et al., 2009; Tawancy, 1996). By providing appropriate slip systems, this work may now contribute to achieving better matching simulation and experimental microstructures. In this context, this study also validates the, however small, yield stress anisotropy, as has been expected and shown to be necessary for appropriate simulations (le Graverend et al., 2011). Further, the effect of slip plane/ growth defect intersections can be incorporated in hardening models with consideration of fracture following plasticity where accumulation of dislocations on individual planes appears likely and there is no or little confinement preventing decohesion along prismatic planes. The results presented here provide a first indication which mechanisms would be most relevant to include in a crystal plasticity model, even if individual components, such as quantitative hardening characteristics and fracture stresses, would have to be determined more accurately by further nanomechanical experiments (Dehm et al., 2018) and correlative simulations and experiments on the relevant microstructures.

In the context of the mechanisms underlying these descriptions of material behaviour, future experiments may build on the results presented here to purposefully investigate the quantitative effect of growth defects on dislocation nucleation and transmission, similar to work elsewhere on nanotwins and grain boundaries in steel and copper (Choi et al., 2017; Imrich et al., 2014), the effect of thermal activation in terms of rate and temperature effects on the plasticity of the μ-phase and the quantitative relationship between defect density on prismatic planes and their cohesive strength. This may be done by pre-deformation and subsequent micro-cantilever compression, although in this case the quantification of the dislocation density on a given plane will present a considerable challenge.

## 5  Conclusions

Using nanoindentation, EBSD- and AFM-assisted surface trace analysis, micropillar compression and TEM, we investigated the mechanical properties and deformation mechanisms in the $Fe_7Mo_6$ μ-phase. The main conclusions are:

- An average hardness and indentation modulus of 11.7$\pm$0.9 GPa and 249$\pm$8 GPa, were measured and basal as well as prismatic planes identified as slip planes using slip trace analysis around indentations.
- Slip was found to take place predominately on basal planes in indentation and at a CRSS of 1.92$\pm$0.22 GPa or 1.79$\pm$0.31 GPa in micropillar compression assuming conventional basal dislocation glide or synchroshear, respectively.
- Prismatic slip was activated where the resolved shear stress on the basal planes was low and the CRSS for the onset of deformation by dislocation glide on the prismatic planes was measured as 2.11$\pm$0.66 GPa by microcompression.
- A strong interaction of growth defects with mobile dislocations was observed with softening where dislocations can move along the pre-existing defects and hardening where the slip plane intersects them.
- Deformation on prismatic planes may concentrate on individual planes, leading to a high local defect density and decohesion of the plane in the absence of confining pressure.

ACKNOWLEDGEMENTS

The authors gratefully acknowledge funding by the German Research Foundation (DFG) within project KO 4603/2-1 and the German Academic Exchange Service (DAAD) within project 57130374. We would also like to thank H. Springer (MPIE, Düsseldorf, D), H.T. Pang (University of Cambridge, UK) and H. Horn-Solle (ILT, Aachen, D) for their help with the sample preparation, annealing and EDX analysis. Further, the authors would like to cordially thank J. S. K.-L. Gibson for his encouragement and constructive criticism on the manuscript.